\documentclass[rnote]{aa}  
\usepackage{graphicx}
\usepackage{txfonts}
\usepackage{color}

\begin{document}

   \title{GTC optical imaging of extremely red 5C radio galaxies at high redshift}

   \subtitle{}

   \author{A. Humphrey\inst{1}\fnmsep\inst{2}, M. Villar-Mart\'{i}n\inst{3} \and P. Lagos\inst{1}
                    }

   \institute{Instituto de Astrof\'{i}sica e Ci\^encias do Espa\c{c}o,
     Universidade do Porto, CAUP, Rua das Estrelas, PT4150-762 Porto, Portugal\\
              \email{andrew.humphrey@astro.up.pt}
         \and
             Instituto Nacional de Astrof\'{i}sica, \'Optica y
             Electr\'onica (INAOE), Aptdo. Postal 51 y 216, 72000 Puebla, Pue., Mexico\\
           \and
           Centro de Astrobiolog\'{i}a (INTA-CSIC), Ctra de Torrej\'on a Ajalvir, km 4, E-28850 Torrej\'on de Ardoz, Madrid, Spain\\
           }

   \date{Accepted for publication in A\&A.}

 
  \abstract
   {}
   {We investigate the nature of seven unusual radio galaxies from the 5C catalogue that were previously known to have extremely red $R-K$ colours, and for which emission lines were previously found to be weak or absent in their optical spectra.}
   {We present and discuss $u$, $g,$ or $r$ images of these
     radio galaxies, obtained using the Optical System for Imaging and low-Intermediate-Resolution Integrated Spectroscopy (OSIRIS) at the Gran
     Telescopio Canarias (GTC).}
   {We have detected all seven targets in our $g$-band imaging. Their optical emission is extended, and we tentatively detect a radio-optical alignment effect in this sample. A subset of our sample (three sources) shows broad-band spectral energy distributions that flatten out near the wavelength range of the $g$-band, implying a dominant contribution there due to young stars and/or scattered
or reprocessed radiation from the active nucleus.}
   {}

   \keywords{}

   \maketitle

\section{Introduction}
The high-redshift Universe offers a goldmine of as yet untapped information about powerful AGN activity and how the AGN interacts with its host galaxy. Among the various classes of high-$z$ galaxies, powerful radio galaxies stand out as among the best probes of the symbiosis between AGN activity and galaxy evolution. These progenitors of present-day massive elliptical galaxies are highly luminous across most of the electromagnetic spectrum, are rich in dust and gas, and are frequently associated with prodigious star formation (SF) activity (see Seymour et al 2007; Miley and De Breuck 2008; Drouart et al. 2014). 

Although most powerful radio galaxies show bright emission lines, it has been found that up to $\sim$30 per cent of all steep-spectrum ($\alpha$$\la$-0.5, where $S_v \propto v^{\alpha}$) high-$z$ ($z\ga$1) radio galaxy candidates do not show emission lines, even after deep spectroscopic observations on medium to large telescopes (see e.g.  Reuland 2005; Miley \& De Breuck 2008). These radio galaxies also commonly have extremely red optical to near-IR colours, implying either heavy reddening or an old stellar population (Willott et al. 2001: hereafter W01), they
often show compact radio sources, and are typically bright at sub-millimetre wavelengths (Reuland 2005). It is tempting to
consider that this class of weak-lined radio galaxy may represent an important but often overlooked phase in AGN host co-evolution that has the potential to improve our understanding of galaxy evolution. We refer to them here as line-dark and red radio galaxies (LDRRGs). 

In this Research Note, we present results of deep optical imaging of seven LDRRGs from the 5C sample of W01 that have spectroscopic or tentative photometric redshifts in the range $z\sim$1-2. By specifically targeting emission in the blue part of the observer-frame optical, we may examine the structural components that emit blueward of the 4000 \AA~break, such as young stars or AGN-powered emission nebulae. We assume $\Omega_{\Lambda}$ = 0.7, $\Omega_M$ = 0.3, $H_0$ = 71 km s$^{-1}$ Mpc$^{-1}$ throughout.

\begin{table*}
\small
\centering
\caption{Journal of GTC imaging observations and sample. The magnitudes colours are from Table 1 of W01 and were measured by those authors in 3\arcsec apertures. They are in the AB system. Redshifts shown in parentheses are tentative values from W01, based on fitting twelve simple stellar populations to their $R$ through $K$ broad-band SEDs, or in the case of 5C 6.62, on their tentative detection of a single emission line thought to be H$\alpha$. The firm redshift determination for 5C 7.245 comes from Humphrey et al. (2015), who detected four emission lines at $z=$1.609, confirming the tentative detection of H$\alpha$ at $z=$1.61 reported
by W01.} 
\begin{tabular}{lllllllllll}
\hline
Target & Alt. name & RA (J2000) & Dec (J2000) & Date & Filter & T (s) & Seeing & z & $K$ & $R-K$ \\

\hline
5C 6.17 & B2 0206+34 & 02h09m20.0s & +34d28m35s & 4/9/2010 & $g$ & 1900 & 0.7\arcsec-0.8\arcsec & (1.05) & 20.36$\pm$0.07 & $>$4.5 \\   
             & & & & 4/9/2010 & $r$ & 540 & 0.7\arcsec-0.8\arcsec & & & \\   
5C 6.62 & B2 0210+32 & 02h13m27.1s & +33d08m03s & 4/9/2010 & $g$ & 1900 & 0.7\arcsec-0.8\arcsec & (1.45) & 19.99$\pm$0.08 & 4.9$\pm$0.5 \\       
5C 6.83 & B2 0211+30B & 02h14m06.7s & +30d53m49s & 5/9/2010 & $g$ & 1900 & 0.8\arcsec-0.9\arcsec & (1.8) & 20.22$\pm$0.08 & $>$4.6 \\
              & & & & 5/9/2010 & $r$ & 360 & 0.8\arcsec-0.9\arcsec & & & \\       
5C 6.242& B2 0218+31 & 02h21m14.4s & +31d17m06s & 5/9/2010 & $g$ & 1900 & 0.8\arcsec-0.9\arcsec & (1.9) & 20.70$\pm$0.10 & $>$5.8\\
               & & & & 5/9/2010 & $r$ & 360 & 0.8\arcsec-0.9\arcsec & & & \\
5C 7.47  & B2 0812+24 & 08h25m59.2s & +25d53m00s & 28/2/2011 & $u$ & 816 & 1.1\arcsec-1.3\arcsec & (1.7) &  21.44$\pm$0.10  & 2.4$\pm$0.2 \\
              & & & & 28/2/2011 & $g$ & 800  & 1.1\arcsec-1.3\arcsec & & & \\
5C 7.208 & B2 0820+25 & 08h23m17.9s & +24d56m40s & 28/2/2011 & $g$ & 800  & 1.0\arcsec & (2.0) & 20.03$\pm$0.08 & 4.0$\pm$0.2 \\    
5C 7.245 & B2 0822+26 & 08h25m57.0s & +26d43m58s & 28/2/2011 & $g$ & 800 & 1.0\arcsec & 1.609 & 20.09$\pm$0.08 & $>$4.3 \\
\hline
\end{tabular}
\label{tab1}
\end{table*}

\begin{figure}
\includegraphics{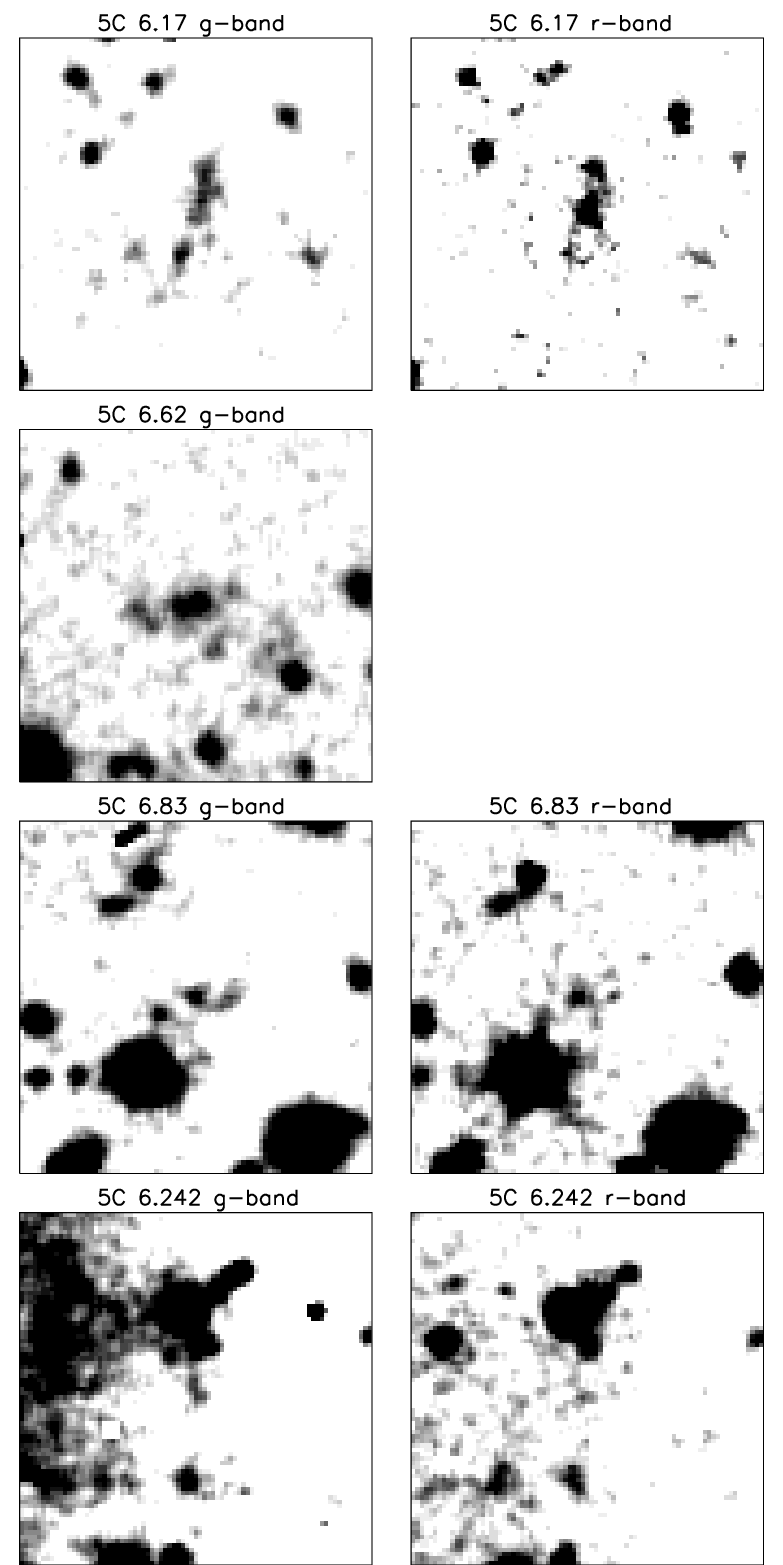}
\includegraphics{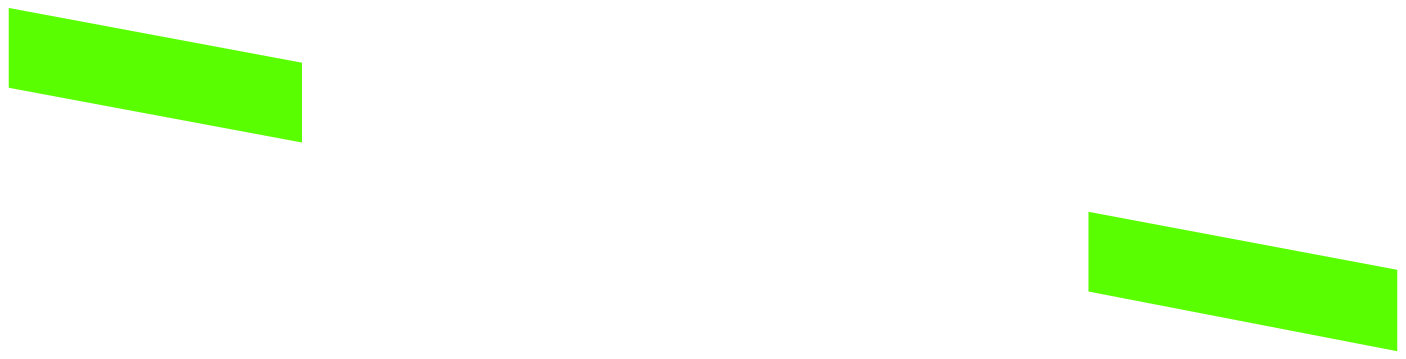}
\includegraphics{fig5C6.17.eps}
\vspace{7.12in}
\caption{Cut-outs from the $g$- and $r$-band images of the 7C-I radio galaxies, showing 20\arcsec$\times$20\arcsec centred on the position of the radio galaxy.  Green bars show the position angle of the radio emission. With the exception of the $r$-band image of 5C 6.17, the images have been smoothed using a 3-pixel boxcar average.  The images of 5C 6.242 show a significant horizontal gradient in the backround emission that is due to ghosts from a bright star positioned elsewhere on the detector array.}
\label{im1}
\end{figure}

\begin{figure}
\includegraphics{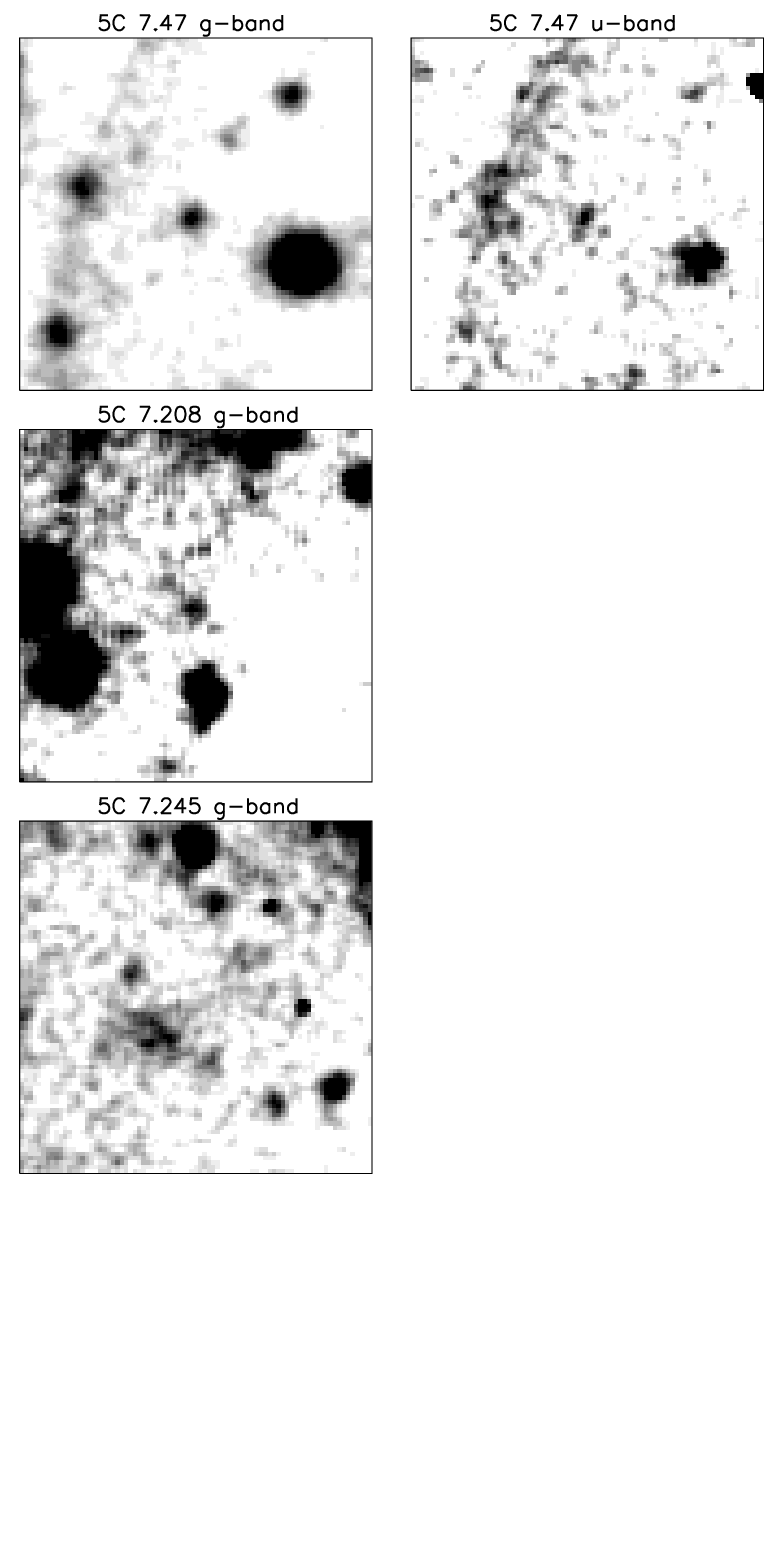}
\includegraphics{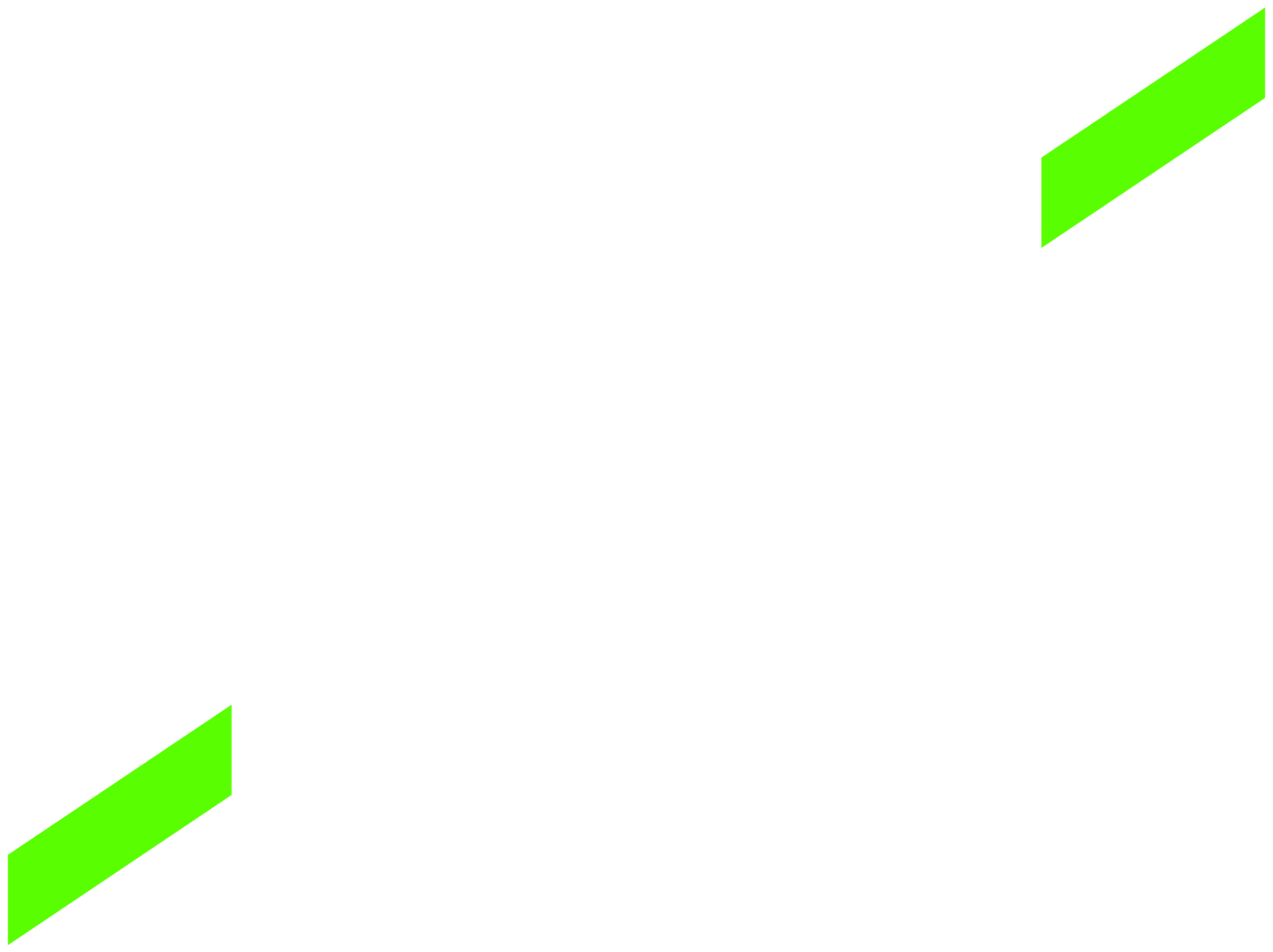}
\includegraphics{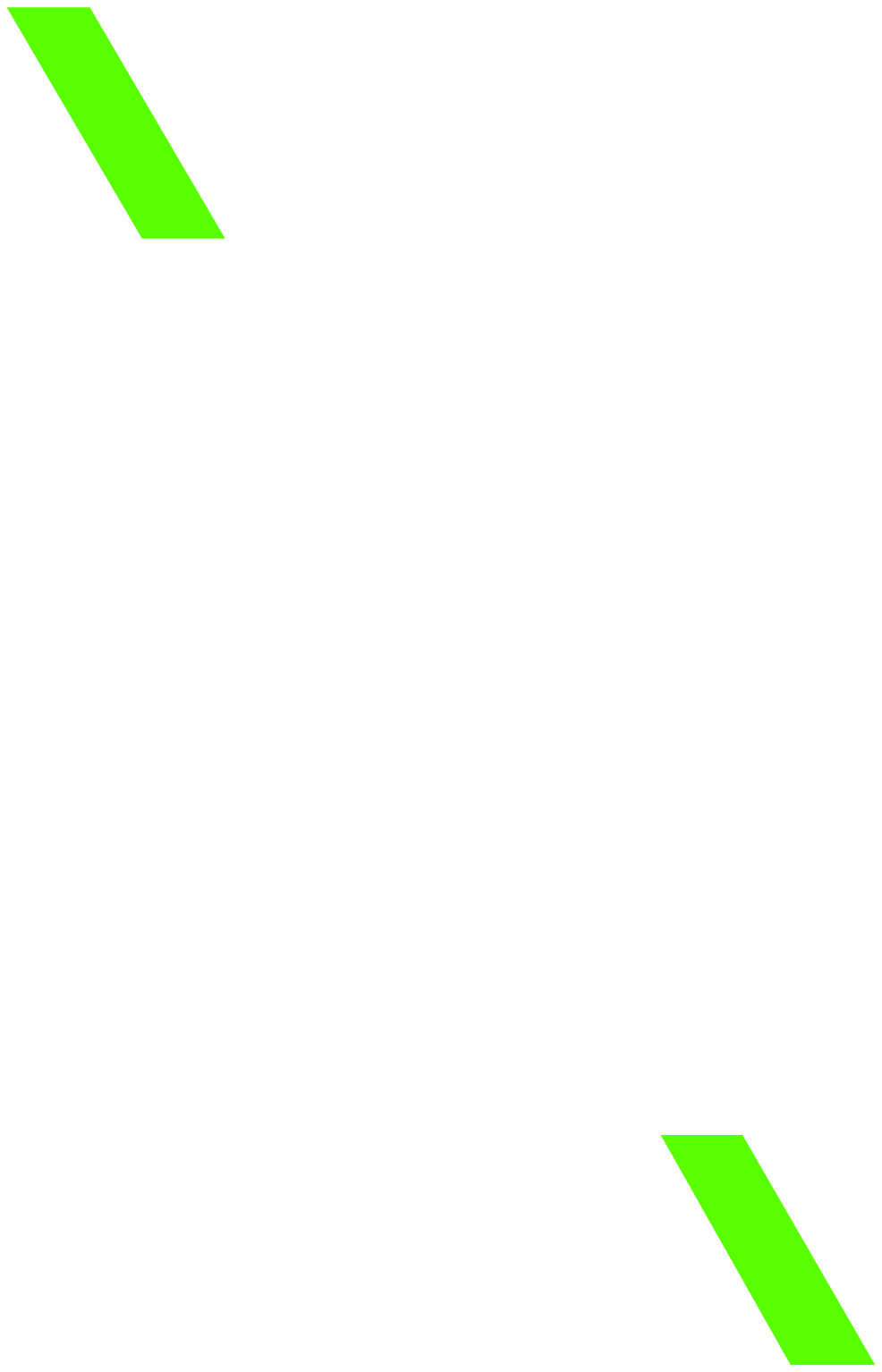}
\vspace{5.3in}
\caption{Cut-outs from the g- and u-band images of the 7C-II radio galaxies, showing 20\arcsec$\times$20\arcsec centred on the position of the radio galaxy. Green bars show the position angle of the radio emission. The images have been smoothed using a 3-pixel boxcar average.  The image of 5C 7.208 is centred on the southern emission component (5C 7.208 S).  The images of 5C 7.47 show ghost-arcs that are due to a bright star positioned elsewhere on the detector array.}
\label{im2}
\end{figure}

\section{Sample and observations}
Our sample is comprised of seven 5C radio galaxies from the 7C-I and 7C-II regions of the 7C redshift survey (e.g. Willott et al. 2002; Willott et al. 2003) that show extremely red optical to near-infrared colours and did not show any strong emission lines in their optical spectra (W01). By fitting simple stellar populations to their $R$ through $K$ broad-band spectral energy distributions (SEDs), W01 obtained tentative photometric redshifts for five of the LDRRGs (5C 6.17, 5C 6.83, 5C 6.242, 5C 7.47,
and 5C 7.208) and obtained tentative spectroscopic redshifts based on single line detections for two further LDRRGs in our sample (5C 6.62 and 5C 7.245). Humphrey et al. (2015) have recently obtained a firm spectroscopic redshift for 5C 7.245 after detecting four rest-frame ultraviolet emission lines. 

The redshifts of all seven LDRRGs in our sample, tentative or otherwise, lie in the range $\sim$1-2 (see Table ~\ref{tab1}). This means that the 4000 \AA~break should lie redward of the upper wavelength cut-off of the Sloan Digital Sky Survey $g$-band filter (5580 \AA) in all cases. As such, observations sample spectral regions blueward of this break, where young stars and AGN activity often make their mark more prominently\footnote{Only at $z<$0.4, rather far from the redshifts (or redshift estimates) of our targets, would the 4000 \AA~ break enter the $g$-band filter bandpass.}. 

The observations were taken in service mode at the Gran Telescopio Canarias (GTC) on La Palma, Spain, during 4 and 5 September 2010 and 28 February 2011 (programme ID: GTC2-10BIACMEX).  The images were taken through Sloan Digital Sky Survey $u$ (3200-3800 \AA), $g$ (4050-5580 \AA), and/or $r$ (5530-7290 \AA) filters, using the Optical System for Imaging and low-Intermediate-Resolution Integrated Spectroscopy (OSIRIS).  Pixel binning of 2$\times$2 was used, resulting in a plate scale of 0.254\arcsec~ per pixel. During these observations, the full-width at half maximum (FWHM) of the seeing disc ranged between 0.7\arcsec and 1.3\arcsec.  The sky was dark, with photometric or clear transparency. See also Table 1. The images were reduced and calibrated in the standard way, making use of software from the IRAF and STARLINK suites (See Humphrey et al. 2015). 

\section{Results}
\begin{figure}
\includegraphics{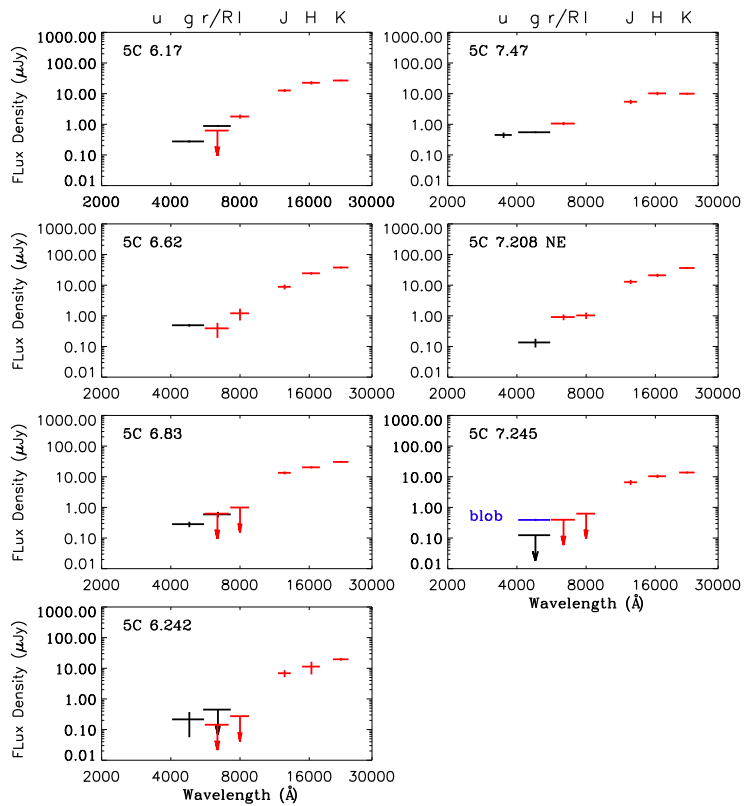}
\vspace{4.1in}
\caption{Broad-band spectral energy distributions of the radio galaxies, showing our new photometry (black) and the photometry of Willott et al. (2001; red).  For 5C 7.245, we also show the $g$-band photometry of the extended blob of emission located $\sim$1\arcsec~ SW of the near-infrared position of the radio galaxy (blue).}
\label{seds}
\end{figure}

Figures ~\ref{im1} and ~\ref{im2} show 20\arcsec$\times$20\arcsec cut-outs from the $u$-, $g$-, and $r$-band images, with the radio galaxy positioned at the centre. Table ~\ref{phot} shows our aperture photometry measurements. Figure ~\ref{seds} shows the optical to near-infrared broad-band SEDs, combining our GTC photometry and the photometry of W01 after converting the latter to AB magnitudes. We used 3\arcsec diameter synthetic apertures centred on the optical emission of the target to match the photometric apertures of W01 as closely as possible. All magnitudes throughout this Research Note are in the AB system. 

{\bf 5C 6.17:} We detected 5C 6.17 in both the $g$- and $r$-band and measured AB magnitudes $r$=24.07$\pm$0.06 and $g$=25.33$\pm$0.08 in a 3\arcsec diameter aperture centred on the $r$-band emission peak of the radio galaxy, which we took to be the position of the hidden active nucleus.  In both filters, the emission is clearly extended to the N and S from the $r$-band emission peak, with a total observed extent of $\sim$6\arcsec (Fig. ~\ref{im1}).  The position angle (north through east) of the extended optical emission is -10$\pm$2$^{\circ}$, in very good agreement with that of the spatially more extended (63\arcsec) NVSS radio emission, which is -10.6$\pm$0.3$^{\circ}$.  This result is reminiscent of the alignment effect in powerful radio galaxies at $z\ge$0.7 (Chambers et al. 1987, McCarthy et al. 1987).   Using a larger aperture of 8\arcsec diameter, which encompasses the entire optical structure of this galaxy, we measured $g$=23.80$\pm$0.06 and $r$=23.45$\pm$0.08, with a relatively blue $g-r$ colour  of 0.4$\pm$0.1, compared to $g-r=1.3\pm$0.1 in the 3\arcsec~ aperture.  While the exact nature of this extended emitting material remains elusive, it seems quite plausible that we have detected a large-scale ($\sim$50 kpc if at $z$=1.05) photoionized and/or reflection nebula powered by continuum radiation from the AGN. 

{\bf 5C 6.62:} The optical counterpart of the radio galaxy is clearly detected in our $g$-band image, with $g$=24.7$\pm$0.1 measured in a 3\arcsec diameter aperture, or $g$=23.58$\pm$0.07 in an aperture of 8\arcsec diameter (Fig. ~\ref{im1}).  Similar to the result of W01, we found the optical emission to be spatially extended.  We measured a total extension of $\sim$6\arcsec in our $g$-band image, with five distinct knots detected within a radius of 4\arcsec. 5C 6.62 also shows a rather blue $g-R$ colour (Fig. ~\ref{seds}), suggesting that star formation and/or nuclear activity might dominate the $g$-band light. The available NVSS radio image is of insufficient resolution to examine the degree of alignment between the radio and UV-optical emitting components. 

{\bf 5C 6.83:} The radio galaxy is detected in both the $g$- and $r$-band images.  It shows a clumpy morphology in the $g$-band image, with three discrete knots within $\sim$5\arcsec of each other, along a position angle of 100$\pm$20$^{\circ}$ (Fig. ~\ref{im1}).  In a 3\arcsec aperture centred on the radio galaxy, we measured $g$=25.3$\pm$0.2 and $r$=24.5$\pm$0.2. The existing NVSS radio image is of insufficient resolution to examine the degree of alignment between the radio and UV-optical emitting components. 

{\bf 5C 6.242:} We detected the radio galaxy in the $g$ band and measured $g$=25.6$\pm$0.6 in a 3\arcsec aperture.  This galaxy is extended along a PA of $\sim$0$^{\circ}$ (Fig. ~\ref{im1}).  In the $r$-band image the radio galaxy counterpart is below our formal 3$\sigma$ detection threshold of $r$=24.8. The available NVSS radio image is of insufficient resolution to examine the degree of alignment between the radio and UV-optical emitting components. 

{\bf 5C 7.47:} The source is detected in both the $u$- and $g$-band images, showing a slight elongation along PA$\sim$160$^{\circ}$ in the $u$ image (Fig. ~\ref{im2}).  The clear absence of a Lyman break between $u$ and $g$ implies $z\le$3.1. Moreover, the flatness of the broad-band SED from $u$ through $R$ (with $u-g$=0.2$\pm$0.2) suggests that this part of the spectrum is dominated by young stars and/or AGN-driven emission processes (Fig. ~\ref{seds}). The available FIRST and NVSS radio images are of insufficient resolution to examine the degree of alignment between the radio and UV-optical emitting components. 

{\bf 5C 7.208:} W01 identified two possible optical to near-IR counterparts to this radio galaxy, both positioned along the major axis of the radio emission. Their similar broad-band SEDs led W01 to suggest that they are at a similar redshift. We detected both sources in our $g$-band image.  The flux ratio between the two sources reverses from the near-infrared to the optical, with the southern source being the brighter of the pair in the $g$-band image, and also the bluer of the two. It is not clear which of the two optical-infrared sources is the host galaxy of the AGN, but W01 have assumed the northern one to be the AGN host galaxy and only showed the broad-band SED of this counterpart. The northern source shows a particularly interesting optical SED (Fig. ~\ref{seds}), having relatively constant flux density across the $R$ and $I$ bands (as noted by W01), then a sharp downward break between $R$ and $g$, suggestive of (i) the Lyman break (consistent with the SED fits of W01); (ii) the 4000 \AA~ break within the $g$
band (which would require z$<$0.4); or (iii) a high equivalent width emission line within the $R$ bandpass. 

{\bf 5C 7.245:} This radio galaxy has been discussed in detail in Humphrey et al. (2015), but is included here for completeness. Our $g$-band image shows no significant emission at the position of the peak of the W01 $K$-band emission, but we detected a blob of extended emission located $\sim$2\arcsec SE of the $K$-band position, along a similar position angle to that of the radio source, with a fainter blob diametrically opposed to the NW (Fig. ~\ref{im2}). Long-slit spectroscopy presented by Humphrey et al. (2015) detected no lines from the $K$-band position at a 3$\sigma$ limit of $\sim$3$\times$10$^{-17}$ erg s$^{-1}$ cm$^{-2}$. However, these authors detected HeII $\lambda$1640, CII] $\lambda$2326, [NeIV] $\lambda$2423 and MgII $\lambda$2800 at the position of the $g$-band blob to the south-east at a redshift of $z=$1.609, consistent with the tentative detection of H$\alpha$ at $z=$1.61
reported by W01. Humphrey et al. (2015) found the emission line ratios of the $g$-band blob to be inconsistent with photoionization by young stars or an AGN, but consistent with ionization by fast shocks. Based on the unusual gas geometry seen in 5C 7.245, Humphrey et al. (2015) argued that we here witness a rare (or rarely observed) phase in the evolution of quasar hosts, when stellar mass assembly, accretion onto the back hole, and/or powerful feedback activity have eradicated cold gas from the central $\sim$20 kpc, but is still in the process of cleansing cold gas from its extended halo. 

\begin{table}
\centering
\caption{Aperture photometry in the AB magnitude system.  Synthetic apertures of 3\arcsec diameter were used.} 
\begin{tabular}{llll}
\hline
Galaxy & $u$            & $g$              & $r$          \\
\hline
5C 6.17 & --           & 25.33$\pm$0.08 & 24.07$\pm$0.06  \\
5C 6.62 & --           & 24.7$\pm$0.1   & --                    \\
5C 6.83 & --           & 25.3$\pm$0.2   & 24.5$\pm$0.2  \\
5C 6.242& --           & 25.6$\pm$0.6   & $\ge$24.8       \\
5C 7.47 & 24.8$\pm$0.2 & 24.58$\pm$0.07   & --        \\
5C 7.208 N  & --           & 26.1$\pm$0.5   & --              \\
5C 7.208 S  & --           & 25.8$\pm$0.3   & --               \\
5C 7.245 K & --           & $\ge$26.2      & --                   \\
5C 7.245 blob & --           & 24.94$\pm$0.09   & --       \\
\hline
\end{tabular}
\label{phot}
\end{table}

\section{Discussion and conclusions}
We have presented broad-band optical images, photometry, and SEDs for seven LDRRG high-$z$ radio galaxy candidates from the 5C / 7C radio surveys. Despite their extremely red $R-K$ colours, all seven have been detected in our $g$-band imaging. 

Furthermore, all seven of our LDRRGs show spatially extended ($\ge$1\arcsec) optical emission structures in the form of a single elongated structure or several aligned knots. Three of our LDRRGs have optical emission structures that are well aligned with the radio-emitting structures (i.e., within 10$^{\circ}$: 5C 6.17, 7.208 and 7.245). Radio images of sufficiently high spatial resolution are currently unavailable for the remaining four LDRRGs in our sample (5C 6.62, 6.83, 6.242, and 7.47), and thus their radio-optical structural alignment cannot be tested. Although this analysis suffers from low number statistics, the results are consistent with the presence of a radio-optical alignment effect in LDRRGs, similar to that found in optically more luminous radio galaxies (Chambers et al. 1987; McCarthy et al. 1987). However, we are unable to ascertain from our data whether the radio-optical alignments in these LDRRGs are due to jet-induced star formation, for instance, or if instead the reprocessing of anisotropically emitted nuclear continuum emission is the
reason. 

Our observations also add new information to complement the existing $RIJHK$ photometry of W01. In several cases (5C 6.62, 6.242, 7.47, 7.245), our data reveal a strong flattening in the blue part of the optical regime ($g-r$, $g-R$ or $u-g$ $<$1.0), consistent with young stars and/or with reprocessed AGN emission dominating this spectral range. These colours, however, would be inconsistent with an evolved  ($\ga$1 Gyr) stellar population -- the colours would be $>$1.0 at moderate to high redshift, based on our redshifting of Bruzual \& Charlot (2003) stellar population spectra. However, for our other three LDRRGs, the SED remains rather red from $K$ through to $g$ ($g-r$ or $g-R >$1), consistent with strong reddening by dust and/or the dominance of evolved stars. 

Finally, we comment that the relatively blue spectral components within the SED of some LDRRGs may potentially affect photometric redshift estimates by diluting (or hiding) spectral features such as the 4000 \AA~ break. Detailed modelling, which is beyond the scope of this Research Note, is needed to properly quantify this effect.

\begin{acknowledgements}
The authors thank Leandro Cardoso, Jean Michel Gomes, Breezy Oca\~na Flaquer, Itziar Aretxaga, Catarina Lobo, Mercedes Filho, Luc Binette, and Bjorn Emonts for useful suggestions and discussions concerning various topics relevent to this work. AH and PL acknowledge Funda\c{c}\~{a}o para a Ci\^{e}ncia e a Tecnologia (FCT) support through UID/FIS/04434/2013, and through project FCOMP-01-0124-FEDER-029170 (Reference FCT PTDC/FIS-AST/3214/2012) funded by FCT-MEC (PIDDAC) and FEDER (COMPETE), in addition to FP7 project PIRSES-GA-2013-612701. AH also acknowledges a Marie Curie Fellowship co-funded by the FP7 and the FCT (DFRH/WIIA/57/2011) and FP7 / FCT Complementary Support grant SFRH/BI/52155/2013. The work by MVM has been funded with support from the Spanish Ministerio de Econom\'{i}a y Competitividad through the grant AYA2012-32295. PL is supported by a post-doctoral grant SFRH/BPD/72308/2010, funded by the FCT. We also thank the anonymous referee for suggestions that helped to improve this Research Note.

\end{acknowledgements}

\end{document}